\documentclass{aastex631}
\usepackage{etoolbox}

\begin{document}

\title{GRB 221009A: A light dark matter burst or an extremely bright Inverse Compton component?}

\correspondingauthor{M. M. González}
\email{magda@astro.unam.mx}

\author[0000-0002-5209-5641]{M. M. González}
\affiliation{Instituto de Astronomía, Universidad Nacional Autónoma de México, Ciudad de Mexico, 04510, Mexico}

\author[0000-0002-4020-4142]{D. Avila Rojas}
\affiliation{Instituto de Física, Universidad Nacional Autónoma de México, Ciudad de Mexico, 04510, Mexico}

\author[0000-0002-8940-5316]{A. Pratts}
\affiliation{Instituto de Física, Universidad Nacional Autónoma de México, Ciudad de Mexico, 04510, Mexico}

\author[0000-0002-2565-8365]{S. Hernández-Cadena}
\affiliation{Instituto de Física, Universidad Nacional Autónoma de México, Ciudad de Mexico, 04510, Mexico}

\author[0000-0002-0173-6453]{N. Fraija}
\affiliation{Instituto de Astronomía, Universidad Nacional Autónoma de México, Ciudad de Mexico, 04510, Mexico}

\author{R. Alfaro}
\affiliation{Instituto de Física, Universidad Nacional Autónoma de México, Ciudad de Mexico, 04510, Mexico}

\author[0000-0002-8774-8147]{Y. Pérez Araujo}
\affiliation{Instituto de Astronomía, Universidad Nacional Autónoma de México, Ciudad de Mexico, 04510, Mexico}

\author[0000-0001-6026-7338]{J. A. Montes}
\affiliation{Instituto de Astronomía, Universidad Nacional Autónoma de México, Ciudad de Mexico, 04510, Mexico}

% Note that the \and command from previous versions of AASTeX is now


\begin{abstract}
Gamma-ray bursts (GRBs) have been considered as potential very high-energy photon emitters due to the large amount of energy released as well as the strong magnetic fields involved in their jets. However, the detection of TeV photons is not expected from bursts beyond a redshift of $z\gtrsim 0.1$ due to their attenuation with the extragalactic background light (EBL). For these reasons, the recent observation of photons with energies of 18 and 251 TeV from GRB 221009A (z=0.151) last October 9th, 2022 has challenged what we know about the TeV-emission mechanisms and the extragalactic background. In order to explain the TeV observations, recent works exploring candidates of dark matter have started to appear.
In this paper, we discuss the required conditions and limitations within the most plausible scenario, synchrotron-self Compton (SSC) radiation in the GRB afterglow, to interpret the one 18-TeV photon observation besides the EBL. To avoid the Klein-Nishina effect, we find an improbable value of the microphysical magnetic parameter below $10^{-6}$ for a circumburst medium  value $> 1{\rm cm^{-3}}$ (expected in the collapsar scenario). Therefore, we explore possible scenarios in terms of ALPs and dark photon mechanisms to interpret this highly-energetic photon and we discuss the implications in the GRB energetics. We find that the ALPs and dark photon scenarios can explain the 18 TeV photon but not the 251 TeV photon. 

\end{abstract}



\section{Introduction} \label{sec:intro}

Gamma-ray bursts (GRBs) are characterized by a prompt emission in the energy range of keV to MeV with a wide range of spectral and temporal features. They also present a late phase called afterglow which is observed at energies from radio to $\gamma$-rays with longer duration than the prompt emission. They are classified based on their hardness ratio \citep{2000PASJ...52..759Q,2001A&A...369..537Q} and duration of their prompt emission \citep{1993ApJ...413L.101K}. In particular, longer bursts are those with a duration larger than 2 s and are associated with the death of massive stars through core-collapse \citep{1993ApJ...405..273W, 1999ApJ...524..262M}. The fireball model \citep{1986ApJ...308L..43P,1986ApJ...308L..47G,1978MNRAS.183..359C} explains the prompt emission as the dissipation of kinetic energy in internal shocks and the afterglow phase through shocks generated by the collision of the expanding blast wave with the external medium \citep[see ][ for a complete review]{2004RvMP...76.1143P}. In both cases, synchrotron radiation is the responsible cooling mechanism only up to the energies of a few GeVs \citep{2013ApJ...771L..33W, 2010MNRAS.409..226K, 2012ApJ...755...12V}. Thus, high ($> 10$ GeV) and very-high-energy (VHE, $> 100$ GeV) emissions are described (for high emission) and predicted (for VHE emission) by inverse Compton scattering of lower energy photons by electrons \citep{2004ApJ...604L..85A,2001ApJ...546L..33W} at different regions of the jet and by other hadronic models that we do not discuss here because of the known limited baryonic content of the jet \citep{2004A&A...418L...5D}. In leptonic models, the VHE emission appears delayed with respect to the prompt emission as a consequence of the time for the shock to approach the deceleration radius \citep{1994MNRAS.269L..41M} and the time that takes the $\gamma\gamma$-opacity to decrease \citep[e.g., see][] {2006ApJ...650.1004B}. In fact, the striking detections of GRB 180720B, GRB 190829A and GRB 190114C above energies of 100, 200, and 300 GeV by the H.E.S.S. and MAGIC observatories \citep{2021Sci...372.1081H,2019Natur.575..464A} were described as inverse Compton emission. In all cases, the emission appeared hours after the prompt phase.

On October 9th, 2022, one of the most luminous (or probably the most luminous depending on the final corrected flux) gamma-ray bursts ever recorded was detected by the Fermi-GBM instrument \citep{2022GCN.32636....1V}, followed by SWIFT \citep{2022GCN.32632....1D} and by several other missions as well. Initial observations by GBM show that GRB 221009A consists of two main emission episodes followed by a long tail; the first one with a duration in the tens of seconds and a fluence of approximately $2.2\times 10^{-5}\; \mathrm{erg} \;\mathrm{cm}^{-2}$. The second, a multi-peaked emission, lasting around 327 s (\cite{2022GCN.32642....1L}, preliminary T90) reached a fluence of approximately $2.912 \times 10^{-2}\; \mathrm{erg} \;\mathrm{cm}^{-2}$ \citep{2022GCN.32642....1L}. The duration of the tail is still being determined but it is in the order of two thousand seconds. While other missions detected GRB 221009A at relatively the same time periods, all of them reported different estimations for the fluence that are being revised due to the saturation of the instruments. Nevertheless, all the fluences reported for the second emission episode remain in the order of $10^{-2}\; \mathrm{erg} \;\mathrm{cm}^{-2}$. The ground-based locations are consistent with uncertainties. Observations of the counterpart began at approximately T0+150 s in the X-ray band \citep{2022GCN.32671....1T} and several hours later in the IR, radio, and optical bands \citep{2022GCN.32680....1T,2022GCN.32852....1P,2022GCN.32644....1H,2022GCN.32655....1F, 2022GCN.32767....1L,2022GCN.32654....1D,2022GCN.32652....1B}. Moreover, spectroscopic observations of the afterglow with the GTC's OSIRIS and the VLT's X-Shooter instruments resulted in an estimated redshift of $z=0.151$ \citep{2022GCN.32686....1C, 2022GCN.32648....1D}.

Among the various missions that observed GRB 221009A, the LHAASO observatory reported a detection with two of its three sub-arrays; WCDA and KM2A. LHAASO-WCDA array detection was centered at RA = 288.3$^{o}$, DEC = 19.8$^{o}$, within 2,000 s after the trigger time,  with energies above 500 GeV and significance above $100 \sigma$, while LHAASO-KM2A's detection had a significance above $10 \sigma$ and the highest energy photon at 18 TeV. This marks the first-ever detection \citep{2022GCN.32677....1H} of a GRB with energies above 10 TeV. The detection of an 18 TeV photon is certainly already intriguing, but if true, the report by the Carpet-2 air shower array of a photon-like air shower that corresponds to a 251 TeV photon \citep{ATel15669} is amazing considering its redshift of $z=0.151$. However, \citet{2022ATel15675....1F} have already identified some PeV sources with locations consistent with the position of the 251 TeV photon. 

Given what has been mentioned above regarding the observations of the brightest transient event ever seen, a fast conclusion could mislead the description of GRB 221009A toward inverse Compton scattering. However, it is the need to understand how to avoid the EBL attenuation of a flux of photons with energies of 18 TeV (in the best scenarios the attenuation factor is $\sim \exp (-6)$  \citep{2008A&A...487..837F} and 251 TeV (attenuation factor is $\sim \exp (-77)$) that takes us to consider alternatives that invoke Dark Matter (DM) particles. Some authors \cite{2022arXiv221013349D} have approached the problem through the acceleration of hadronic particles in the jet since it has the ability to accelerate extremely high-energy cosmic rays, and their dispersion causes an electromagnetic cascade in ISM. The TeV emission detected by LHAASO could be explained by the line of sight component of this flux. Some studies also explore Lorentz Invariance violation effects on  $\gamma - \gamma$  absorption \citep{2022arXiv221006338L, 2022arXiv221011261F,2022arXiv221101836Z}.

The nature of DM is still unknown, although according to the best cosmological model ($\Lambda CDM$ ), Dark Matter constitutes around 25 \% of the energy density of the universe \citep{2020}. There are several candidates that have been proposed that could describe the nature of dark matter. Among them, the most studied candidate are Weakly Interactive Massive Particles (WIMPs) \citep{Roszkowski_2018} which arrive naturally from a super-symmetric extension of the standard model. However, the Large Hadron Collider at CERN has not found evidence of supersymmetric particles, and other DM candidates are being considered, such as Axion-Like particles (ALPs) and dark photons \citep{Beskidt_2012}. 

Dark photons are proposed to be the gauge bosons of a dark sector under the group $U(1)$ \citep{HOLDOM1986196}. They are considered to be massive and predominantly interact with photons from the standard model.
A kinetic mixing/coupling with the standard model (SM) photon is possible, leading to photon-dark photon conversions either under the presence of magnetic fields or in the vacuum. Because of the small interactions between dark photons and other SM particles, it is possible for dark photons to travel cosmological distances.

ALPs are pseudo-scalar bosons that generalize the concept of Quantum Chromodynamics (QCD) Axions and arise from models Beyond the Standard Model (BSM). Their production happened in the early inflationary Universe. Both of them, Axions and ALPs, are excellent candidates to be the dark matter of the Universe \citep{ABBOTT1983133, PRESKILL1983127, DINE1983137} . Since they are cold  dark matter (CDM), they agree with $\Lambda$CDM \citep{Visinelli_2010}.

Currently, DM and gamma-ray bursts have been linked only by a few authors. The observations of photons with energy of tens to hundreds of GeV are explained as a possible signature of DM; even though, Inverse Compton scattering is considered to be the most plausible mechanism. However, the combination of VHE emissions and a moderate redshift in GRB 221009A cannot be easily explained by just Inverse Compton scattering, and more complex scenarios are needed. Among different possibilities, we summarize three scenarios of DM production in GRBs that could fit some of the features observed in GRB 221009A.
 
 Scenario one: Several studies \citep{alpSNI,alpSNII} proposed the production of light DM candidates during the core collapse of a massive star, leading to the production of a burst of DM that may escape from the surface of the star. We review the case for ALPs, but the same conclusions are valid for dark photons. For ALPs, different mechanisms contribute to the total production rate such as the Primakoff process, Bremsstrahlung, electron-positron fusion, and photon coalescence. The dominant contributions to the production rate depend on the mass of the ALP, see \citep{alpSNII}. However, due to screening effects and the kinematic mass of the photon, the maximum energy estimated for the ALPs produced during the core collapse is lower than $1~\text{GeV}$. This means that the probability to generate ALPs with energies $E>~\text{TeV}$ is very low. There are searches for ALPs under these hypotheses by using data from the Gamma-Ray Spectrometer \citep{SN1987AI} and Fermi-LAT Observatory \citep{SN1987AII, SNFermi} within energies below $300~\text{MeV}$. If GRB 221009A is likely the result of the core collapse of a massive star, (\cite{2022GCN.32686....1C}, \cite{amati2022} further studies are needed to extend the production of ALPs or any other dark-matter particle to energies of at least to tens of TeV. A BdHNI typical binary-driven hypernova type I, carbon-oxygen core with a companion neutron star \citep{2022GCN.32780....1A} is also proposed as the progenitor of GRB 221009A. If this is the case, production of ALPs is still possible in the neutron star, see for example \citep{ALPsNeutronMergers}. The neutron star survives some time after the burst allowing the production of ALPs \citep{SN1987AII}.

A second possible scenario to create light DM is the fusion of pairs of electron-positron ($e^{+}~e^{-}$ fusion) to light DM, \citep{alpSNII}. This process should be able to occur with particles produced after the burst flash, and even during the afterglow. The maximum energy of the ALPs produced in this case must be at least the observed photon energy of 18 TeV (or even the 251 TeV photon). 

Finally, the third scenario implies that the DM in the Universe could be multi-component, including a heavy component (for example, a WIMP with masses above $100~\text{TeV}$) plus a light DM component (either an ALP or dark photon candidate). Scenarios with multi-component dark matter have been explored in colliders and direct and indirect DM searches, for example, \citet{MDMI, MDMII, MDMIII}. In this case, during the core collapse of the progenitor (GRB 221009A), the WIMP annihilation could be enhanced by particle capture from the star \citep{DMSNI} and  produce photons with a continuous spectrum at energies near the mass of the candidate. This emission should be isotropic, and the photons produced during the WIMP annihilations are converted to ALPs and converted back to photons after entering the Milky Way to reach observatories on Earth. A variation of this situation is that the DM candidate annihilates directly to dark photons \citep{POSPELOV2009391} that oscillate to standard model photons, which could be detected on the Earth.

\begin{table}[ht]
\centering
\begin{tabular}{ccc}
\hline
Parameter & Value & Reference\\\hline
$E_{\rm iso}$ & $2\times10^{54}$ erg &  \citep{2018ApJ...869L..23T}\\
$D_{L}$ & $2.2317\times10^{27}$ cm &\\
$\rm p$ & -2.4 & \citep{2015PhR...561....1K}\\
EBL model & Gilmore& \citep{Gilmore_2012}\\
$B_{MW}$  &  $\sim 3 \mu   $G & \citep{Jansson_2012} \\
$B_{HostGal}$ & $\sim 3 \mu  $G & assumed as MW \\
$B_{jet}$  & $10^{6}  $G &  \citep{Mena_2011} \\
$d_{jet}$  & $10^{10} $cm  &  \citep{Mena_2011}\\
$d_{HostGal}$ host & 30 $\rm kpc$ & assumed as MW\\
$d_{MW}$ & $\rm < 30 kpc$ & less than MW diameter\\
$\alpha_{DM}$ & -1.8 & assumed\\
Energy carried by DM & $1-10\%$ & assumed\\ \hline
\end{tabular}
\caption{Parameters considered in our calculations.}
\label{tab:param}
\end{table}

In this paper, we explore these scenarios to explain the high energy emission detected in GRB 221009A.\
First, in Section \ref{sec:IC}, we discussed the values of the microphysical parameters to obtain electron energies below the Klein-Nishina limit so that the Synchrotron Self-Compton (SSC) emission can take place. We assume, as for all long bursts, that the density of the surrounding medium, $\eta$, is larger than 1 $\rm cm^{-3}$. In Section \ref{sec:DM}, we explore an alternative scenario in which there is a release of dark matter by the burst. We estimate the minimum survival probability for a photon to come from a DM oscillation and reach Earth. Then, we explore the candidate's region of ALPs and dark photons as the DM particles generated in the burst.
Finally, we summarize our conclusions and give final remarks.

%%%%%%%%%%%%%%%%%%%%%%%%%%%%%%%%%%%%%%%%%%%%%%%%%%%%%%%%%%%%%%%%%%%%%%%%%%%%%%%%%%%%%%%%%%%%%%%%%%%%%%%%%%
%%%%%%%%%%%%%%%%%%%%%%%%%%%%%%%%%%%%%%%%%%%%%%%%%%%%%%%%%%%%%%%%%%%%%%%%%%%%%%%%%%%%%%%%%%%%%%%%%%%%%%%%%%
\section{Leptonic Inverse Compton Scattering} \label{sec:IC}

Within the leptonic emission models, SSC is the preferred mechanism either in the reverse \citep{2007MNRAS.381..732G, 2007ApJ...655..391K, 1999MNRAS.306L..39M, 2004MNRAS.353..647N} or forward external shocks \citep{2007MNRAS.379..331P, 1998ApJ...497L..17S}. The duration of the emission generated in the reverse shock is expected to be shorter (as flashes) and at energies lower than the emission released in the forward shocks. The two main episodes of GRB 221009A appeared during the first $\sim$ 600 s of the bursts. Although LHAASO does not report the detection time of the 18-TeV photon or whether it is unique or one of many 10 TeV-photons, the duration of the observation is given as $\sim$ 2000 s. Thus, we will assume for the purposes of this analysis, that the VHE emission lasted as long as the LHAASO observation duration which is much longer than the prompt duration of 327 s (without considering the long tail as reported by \citep{2022GCN.32642....1L}). Then, here we will analyze the case of SSC in forward shocks to explain the TeV-photons of GRB 221009A.

The SSC process occurs when the same electron population that radiates synchrotron photons up-scatters them to higher energies as $h\nu^{\rm ssc}_{\rm k}\sim \gamma^2_{\rm k} h\nu^{\rm syn}_{\rm k}$, where the notation ${\rm k=m, c}$ represents the minimum and cooling frequencies and $h$ stands for the Planck constant. The maximum flux that the SSC process can reach $F^{\rm ssc}_{\rm max}\sim k\tau F^{\rm syn}_{\rm max}$ depends on the maximum synchrotron flux and the optical depth.  The spectral breaks and the maximum flux for SSC emission in the afterglow can be expressed as \citep{2001ApJ...548..787S}

{\small
\begin{eqnarray}\label{ssc_br-h}
h\nu^{\rm ssc}_{\rm m}&\simeq& 0.6\,{\rm GeV}\,  \left(\frac{1+z}{2.15} \right)^{\frac54}\,\epsilon_{\rm e,-1}^{4}\,\epsilon_{\rm B,-3}^{\frac12}\,n^{-\frac14}\,E^{\frac34}_{\rm 54}\,t^{-\frac94}_3,\cr
h\nu^{\rm ssc}_{\rm c}&\simeq& 3.9\,{\rm MeV} \left(\frac{1+z}{2.15}\right)^{-\frac34}\left(1+Y \right)^{-4}\epsilon_{\rm B,-3}^{-\frac72}\,n^{-\frac94}\,E^{-\frac54}_{\rm 54}\,t^{-\frac14}_3\,,\cr
F^{\rm ssc}_{\rm max}&\simeq& 7.9\times 10^{-3}\,{\rm mJy}\left(\frac{1+z}{2.15}\right)^{\frac34}\,\epsilon_{\rm B,-3}^{\frac12}\,n^{\frac54}\,d^{-2}_{\rm z, 28}\,E^{\frac54}_{\rm 54}\,t^{\frac14}_3,\,\,\,\,\,\,\,\,
\end{eqnarray}
}

where $Y$ is the Compton parameter, $z=0.151$ is the redshift, $E$ is the equivalent kinetic energy, $d_{z}$ is the luminosity distance calculated by considering the cosmological parameters presented in \cite{Bennett..2014}, $n=1\,{\rm cm^{-3}}$ is the density of the surrounding medium, $\epsilon_{e}$ and $\epsilon_{B}$ are the microphysical parameters related to 
the total energy given to accelerate electrons and to amplify the magnetic field, respectively, with the constraint of  $\epsilon_{e} + \epsilon_{B} < 1$ \citep{2001ApJ...548..787S}.   We adopt the convention ${\rm Q_{x}=\frac{Q}{10^x}}$ in cgs units.  The SSC light curve in the fast-cooling regime is

{\footnotesize
\begin{eqnarray}
\label{ssc_ism1}
F^{\rm ssc}_{\nu}=  \cases{ 
2.6\times 10^{-7}\,{\rm mJy}  \left(\frac{1+z}{2.15} \right)^{\frac38}\epsilon_{\rm B,-3}^{-\frac54}\,n^{\frac18} \,E^{\frac{5}{8}}_{54}\,d^{-2}_{\rm z, 28}t^{\frac18}_3 (h \nu)_{13}^{-\frac12}, \hspace{4cm} \nu^{\rm ssc}_{\rm m}<\nu <\nu^{\rm ssc}_{\rm c},\hspace{.25cm}\cr
2.0\times 10^{-9}\,{\rm mJy} \left(\frac{1+z}{2.15} \right)^{\frac{5p-2}{8}}\,\left(1+Y\right)^{-2}\, \epsilon_{\rm B,-3}^{\frac{p-6}{8}}\,\epsilon_{e,-1}^{2p-2}\,E^{\frac{3p+2}{8}}_{54}t^{-\frac{9p-10}{8}}_3\,(h \nu)_{13}^{-\frac{p}{2}},\,\,\,\, \hspace{1.1cm}  \nu^{\rm ssc}_{\rm c} <\nu\,, \cr
}
\end{eqnarray}
}

and the slow-cooling regime is given by

{\footnotesize
\begin{eqnarray}
\label{ssc_ism2}
F^{\rm ssc}_{\nu}=  \cases{ 
8.6\times 10^{-6}\,{\rm mJy}  \left( \frac{1+z}{2.15} \right)^{\frac{5p+1}{8}}\epsilon_{\rm B,-3}^{\frac{p+1}{4}}\,\epsilon_{e,-1}^{2(p-1)}\,n^{\frac{11-p}{8}} \,d^{-2}_{\rm z, 28}\,E^{\frac{3p+7}{8}}_{54}\,t^{-\frac{9p-11}{8}}_3 (h \nu)_{13}^{-\frac{p-1}{2}}, \hspace{1cm} \nu^{\rm ssc}_{\rm m}<\nu <\nu^{\rm ssc}_{\rm c},\hspace{.25cm}\cr
2.0\times 10^{-9}\,{\rm mJy} \left(\frac{1+z}{2.15} \right)^{\frac{5p-2}{8}}\,\left(1+Y\right)^{-2}\, \epsilon_{\rm B,-3}^{\frac{p-6}{8}}\,\epsilon_{e,-1}^{2p-2}\,E^{\frac{3p+2}{8}}_{54}t^{-\frac{9p-10}{8}}_3\,(h \nu)_{13}^{-\frac{p}{2}},\,\,\,\, \hspace{1.4cm}  \nu^{\rm ssc}_{\rm c} <\nu\,, \cr
}
\end{eqnarray}
}
  
where $(h \nu)_{13}=10\,{\rm TeV}$ and $t_3=10^3\,{\rm s}$ correspond to the energy band and timescale of this process.    A direct effect on the SSC spectrum due to the Klein Nishina regime is the suppression of up-scattered synchrotron photons. Considering the synchrotron afterglow model, the break energy in the KN regime is given by

\begin{equation}
h\nu^{\rm KN}_{\rm c}\simeq 67.2\, {\rm GeV} \left(\frac{1+z}{2.15}\right)^{-\frac34}\, (1+Y)^{-1} \,\epsilon_{B,-3}^{-1}\,n^{-\frac34}\,E^{-\frac14}_{54}\,t^{-\frac14}_3\,,
\end{equation}

\begin{figure}[ht]
        \centering
        \includegraphics[height=0.35\textheight,width=0.60\textwidth]{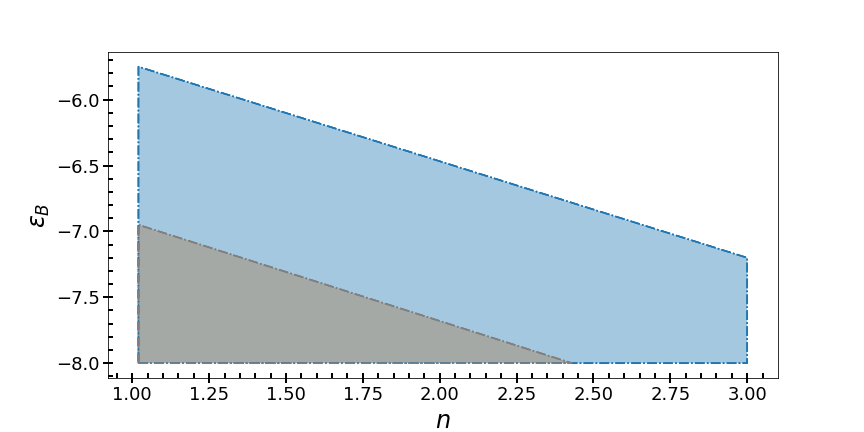}
        \caption{$\epsilon_B$ as a function of the density of the surrounding medium. The blue zone corresponds to an $h\nu^{\rm KN}_{\rm \gamma}>$ 18 TeV and $t=2000$ s while the superimposed grey zone corresponds to $h\nu^{\rm KN}_{\rm \gamma}>$ 251 TeV and $t=5000$ s.  
        \label{fig:Beta}}
\end{figure}

Clearly, the observation of 18 and 251 TeV-photons requires even higher values for $h\nu^{\rm KN}_{\rm \gamma}$. Figure \ref{fig:Beta} shows the allowed values of $\epsilon_B$ assuming $n > 1 {\rm cm^{-3}}$ \citep[typical value for long bursts;][]{2011A&A...526A..23S}, a kinetic energy efficiency of 20\%  \citep{2015PhR...561....1K} and values for the Klein Nishina break energy above 18 TeV at $t=2000$ s and 251 TeV at $t= 5000$ s (See Table \ref{tab:param} for other GRB parameters considered). The choice of $t$ assures that the TeV-photons could have been observed at any time within the observational period. As shown, the required values of $\epsilon_B$ are less than $10^{-6.5}$ or $10^{-7.0}$ depending on the energy of the TeV-photon observed and the observational duration. Some authors \citep{2014ApJ...785...29S} have considered values as low as the ones found here to describe some afterglow observations at low frequencies, but they are not commonly used values. If the TeV-photons come in the first 10 seconds of the burst, these upper limit values are relaxed slightly, but not enough, to values of -5.5 and -6.6 for 18 and 251 TeV Klein Nishina break energies, respectively. 

%%===================
Considering the dependence on time of the SSC lightcurve (Equations \ref{ssc_ism1} and \ref{ssc_ism2}), the maximum flux will happen in the first seconds of the burst. Then, we estimate the flux at the observation times of 1, and 5000 seconds, requiring $h\nu^{\rm KN}_{\rm \gamma}$ greater than 250 TeV (for t=0-5000 s), an $n = 1.2\,{\rm cm^{-3}}$, $\epsilon_{B} = 10^{-7.5}$ and  $\epsilon_{e} = 10^{-6}$. The resulting flux at the energy of 18 TeV decreases from $8 \times 10^{-20}$ $\rm erg\,cm^{-2} s^{-1}$ at t=1 s to $2.1 \times 10^{-23}$ $\rm erg\, cm^{-2} s^{-1}$ at t=5000 s. This flux is already too small to reach the Earth and to be detected by LHAASO even if EBL does not attenuate it. Note that the energy uncertainty of LHAASO allows the energy of the photon to be up to 25.2 TeV where the EBL attenuation is much larger, $\sim \exp(-10)$.

In summary, SSC mechanism in external forward shocks can produce TeV-photons only when extreme and unlikely values of the microphysical parameters are considered in order to avoid a rapid decrement of the cross-section (and the SSC emission) when approaching the Klein-Nishina regime.
Furthermore, even if SSC were the case, the energy flux would be too small to explain the observations of TeV photons as we will discuss in Section \ref{sec:lhaaso}. Then, in the next section, we consider the circumstances under which the emission of DM could be responsible for TeV-photons.

%%%%%%%%%%%%%%%%%%%%%%%%%%%%%%%%%%%%%%%%%%%%%%%%%%%%%%%%%%%%%%%%%%%%%%%%%%%%%%%%%%%%%%%%%%%%%%%%%%%%%%%%%%
%%%%%%%%%%%%%%%%%%%%%%%%%%%%%%%%%%%%%%%%%%%%%%%%%%%%%%%%%%%%%%%%%%%%%%%%%%%%%%%%%%%%%%%%%%%%%%%%%%%%%%%%%%

\section{Dark Matter Scenario}\label{sec:DM}

As discussed in the introduction, the production of DM in GRBs has been studied for energies of a few GeVs. There is still a lot of work to be done before these or similar theories explain the observation of TeV photons in GRBs by the creation of DM at the progenitor stage. Furthermore, the identification of release mechanisms will be an even harder task. Nevertheless, if TeV-photons are observed to be associated with long GRBs at redshifts beyond 0.1, new descriptions using DM are the obvious candidates to avoid the EBL. In this section, we estimate the minimum survival probability for the DM scenarios to explain the observations of GRB 221009A, in particular the one from LHAASO and discuss possible scenarios for ALPs and dark photons candidates. 

\subsection{TeV-Photon Detection}\label{sec:lhaaso}

We assume a fraction of the isotropic energy to be released by the DM candidates with efficiency similar to or lower than the fraction released as kinetic energy of electrons in the SSC framework. We have taken two values, $0.01E_{\rm iso}$ and $0.1E_{\rm iso}$. The mechanisms capable to generate or accelerate DM particles at TeV energies in GRBs are not known today. Hence, we will assume that the energy distribution of DM, as well as the photon flux resulting from the conversion of those DM candidates, can be described by power laws. In particular, the photon flux, $\Phi(E)$, is given by,

\begin{equation}
    \Phi(E) = N_{0}\left(\frac{E}{E_{0}}\right)^{-\alpha}.\label{pl}
\end{equation}

where $\alpha = 1.8$. The isotropic energy is related to the fluence $S$ in an energy range as follows \citep{2018ApJ...869L..23T}:

\begin{equation}
    E_{\rm iso} = \frac{4\pi d_z^{2}}{1+z}S,\label{eiso}
    \label{eqn:eiso}
\end{equation}

 The number of $\gamma$-ray photons observed, $N_{ph}$,  by LHAASO or CARPET-2 in an energy range depends on their effective area $A(E)$ (taken from \citet{Ma_2022} for LHAASO and \citet{2020JETPL.112..753D} for CARPET-2), the duration of the observation, $\Delta t$,  and the survival probability of the photon resulting from a DM oscillation, $P_{\gamma}(E)$.  $N_{ph}$ is given by,

\begin{equation}
    N_{ph} = \Delta t\int\displaylimits_{E_{min}}^{E_{max}}P_{\gamma}(E)A(E)\Phi(E)dE.
    \label{eqn:nph}
\end{equation}

Then, we calculate, with Equations \ref{eqn:eiso} and \ref{eqn:nph}, $N_{0}$, and $P_{\gamma}(E)$, required to describe the observed TeV-photons. In order to simplify the calculation we assume that the survival probability follows a flat distribution in the energy range of interest. This allows us to constrain the minimum value for $P_{\gamma}(E)$ that can explain the $18\;\mathrm{TeV}$ photon measured by LHAASO-KM2A and/or the $251\;\mathrm{TeV}$ photon. We show our results in Table \ref{tab:prob}.

\begin{table}[ht!]
\centering
\begin{tabular}{cccc}
\hline
$E_{\rm iso}$ fraction & Normalization at $1\;\mathrm{TeV}$ & LHAASO 18 TeV & CARPET-2 251 TeV\\
 & [ TeV$^{-1}$ cm$^{-2}$ s$^{-1}$] & Photon Probability  & Photon Probability\\
\hline
$0.01$ & $1.72\times10^{-8}$ & $5.81\times 10^{-5}$ & $> 1$\\
$0.1  $ & $1.72\times10^{-7}$ & $5.81\times 10^{-6}$ & $0.71$\\
\hline
\end{tabular}
\caption{Flux normalization at $1\;\mathrm{TeV}$, $N_{0}$, of the photon flux and the survival probability, $P_{\gamma}(E)$, calculated assuming a fraction of the isotropic energy given to DM to observe at least one photon by LHAASO-KM2A, between energies of $10-25\;\mathrm{TeV}$, and one photon by CARPET-2, between energies of $0.1-10\;\mathrm{PeV}$.}
\label{tab:prob}
\end{table}

The survival probability is low but definitely larger than the attenuation factor by EBL. The photons generated at the GRB are attenuated by the EBL in their journey from the burst to the Earth and this correction is always applied to the photon flux reported by instruments. Moreover, in the case of GRB 221009A the energy of the observed photons, of tens and hundred TeVs, pushes the system to the Klein Nishina regime making it enormously hard to obtain the photon energies and fluxes required to describe the observation. However, if the energy is released as DM, a small fraction, lower than $30\%$, will oscillate into photons before reaching our galaxy. These photons will be subject to EBL attenuation. But, the remaining DM particles will oscillate into photons until reaching the MW resulting in a photon flux in general agreement with the timing and position of the burst. The obtained minimum survival probability of $10^{-6}$ ($10^{-5}$) for a fraction of the $E_{\rm iso}$ energy of $10\%$ ($1\%$), a similar fraction released as kinetic energy of electrons, favors the DM scenario over the SSC in the case of the 18 TeV photon. It is a reasonable value and competitive with respect to the EBL attenuation within the energy range of the photon when considering the LHASSO's energy uncertainty, this is an energy between 10.5 and 25.2 TeV. However, for the 251 TeV photon, the minimum survival probability obtained is close to or even higher than 1 pushing the DM scenario to the exclusion region in the case of ALPs and to dark photon masses below $\sim 2 \times 10^{-6}\, \mu {\rm eV}$ as shown in the following subsection. Thus, the most probable scenario for the 251 TeV-photon is the correlation to close PeV-sources as pointed out by \citet{2022ATel15675....1F}.

%%%%%%%%%%%%%%%%%%%%%%%%%%%%%%%%%%%%%%%%%%%%%%%%%%%%%%%%%%%%%%%%%%%%%%%%%%%%%%%%%%%%%%%%%%%%%%%%%%%%%%%%%%

\subsection{Axion-Like Particles}

Under the effect of magnetic fields, ALPs and photons oscillate into each other. Thus, an ALP from a burst with the kinetic energy of TeVs could oscillate into a TeV-photon at any point of its journey to the Earth, in particular, while entering the Milky Way \citep{Mirizzi_2017}.
 Given the lagrangian of ALPs by \citep{PhysRevD.37.1237}:

\begin{equation}
    \mathcal{L}_{ALP} =\frac{1}{2}(\partial_{\mu}a \partial^{\mu}a - m^{2}_{a}a^{2}) + \frac{1}{4f_{a}}aF_{\mu \nu}\tilde{F}^{\mu \nu}. 
\end{equation}

where $a$ is the ALP field , $m_{a}$ the  ALP mass and $F_{\mu \nu}$ the  Faraday Tensor and $\tilde{F}^{\mu \nu}$ its dual. The coupling between ALPs and electromagnetism is given by:

\begin{equation}
 \mathcal{L}_{a\gamma} = \frac{1}{4f_{a}}aF_{\mu \nu}\tilde{F}^{\mu\nu}= ag_{a \gamma} \vec{E}.\vec{B},
\end{equation}
 
where  $g_{a \gamma}$ is the coupling constant and $\vec{E}$, $\vec{B}$ are the electric and magnetic fields respectively \citep{PhysRevD.37.1237}. 

The ALP-Photon system traveling along the line of sight is given by the propagation equation \citep{PhysRevD.84.105030}. As mentioned earlier, we do not assume a starting population of photons because the preferred mechanism, SSC, can not generate them efficiently. Furthermore, to compensate for the EBL attenuation, the photon flux should increase at least as the EBL attenuation as we require higher energies. Photons could oscillate into DM to avoid EBL but this would also increase the photon flux to compensate for those lost photons that do not oscillate in the host galaxy or in the GRB jet. Then, we consider an initial beam made of just DM given by the density matrix, $\rho = \Phi \Phi ^{\dag}$, at $t=0$ as $\rho(0)= diag (0,0,1)$. Then, the survival probability of a photon, $P_{\gamma}$,  resulting from an ALP oscillation, to be observed on Earth can be obtained from the density matrix and mixing matrix $\mathcal{M} $ that includes all physical parameters such as those for the medium, the magnetic field, electron density, and propagation distance. The evolution equation for the density matrix is given by,

\begin{equation}
    i \frac{\partial \rho}{\partial l}= [\rho , \mathcal{M}],
\end{equation}

where $l$ is the propagation distance for a given medium. The survival probability is then,
\begin{equation}
    P_{\gamma}=\rho_{1,1}(l) + \rho_{2,2} (l) ,
    \label{survalp}
\end{equation} with $\rho_{1,1}$, $\rho_{2,2}$ represent the first and second diagonal elements of the density matrix  \citep{Bi_2021} .

In order to find the survival probability, $P_{\gamma}$, we use the open code \textsc{gammaALP} \citep{Meyer_2021} which allows various astrophysical environments.
We take into account the magnetic fields of the host galaxy, the Milky Way, and the jet of the burst. Clearly, the host galaxy and the jet will trigger the conversion of the released DM into photons too far away from our observational instrument so that the converted photon will be absorbed by the EBL. 
The probability of converting DM to photons outside of the MW is not negligible.  We take it into account as well as the EBL attenuation of such photon. The values of the magnetic fields considered, see Table \ref{tab:param}, are conservative since they have not been reported yet. The host galaxy is assumed to be similar in size and magnetic field to the Milky Way. 

Assuming that the initial ALP beam originates at the core collapse of GRB 221009A, we consider two scenarios. The first one considers the magnetic field of the burst jet, the host galaxy's magnetic field, and lastly the Milky Way's magnetic field. In the second scenario, the jet's magnetic field is not taken into consideration. For the regular component of the Milky Way's magnetic field, we consider the Jansson model. We assumed nominal values of the magnetic field of the jet of $B_{T}=10^{6}G$, $n_{e}=10^{8}cm^{-3}$ , and $l=10^{10} cm$ \citep{Mena_2011}. \textsc{gammaALP} calculates the traveling distance in the Milky Way of a particle given its direction.

\begin{figure}[ht]
        \centering
        \includegraphics[width=1 \textwidth]{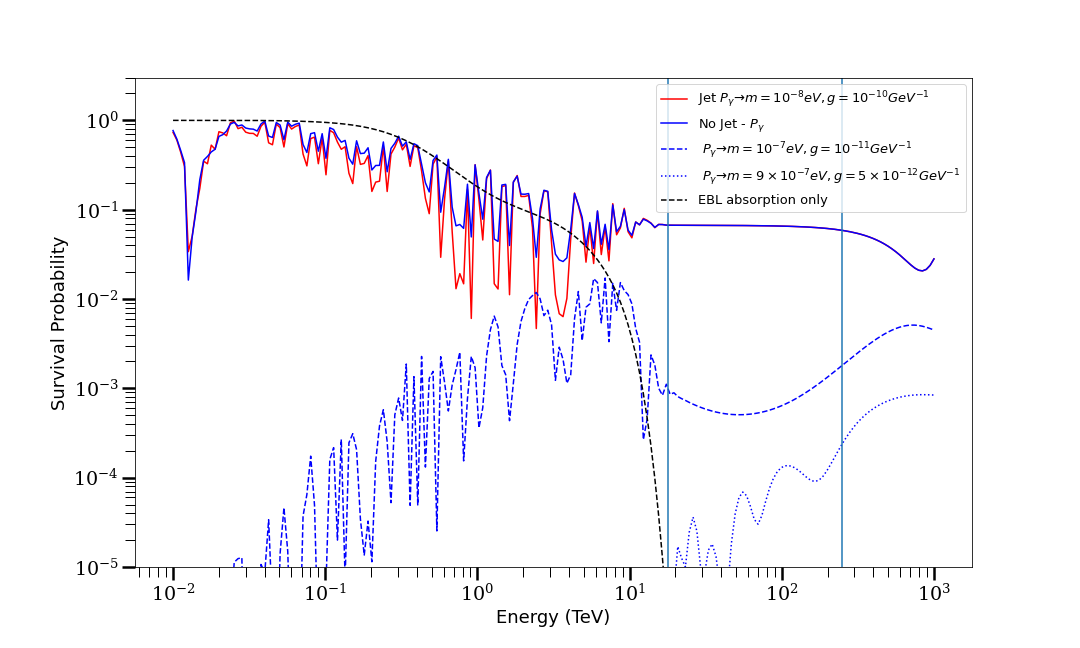}
        \caption{Survival probability of a photon as a function of its energy from an initial beam of ALP candidates defined by m and g.
        The red line considers the magnetic field of the GRB jet while blue lines do not. The vertical lines mark the energies of  18 TeV and 251 TeV as reference.}
        \label{fig:probcases}
\end{figure}

\begin{figure}[ht]\centering   
\epsscale{2}
\includegraphics[width=0.8\textwidth]{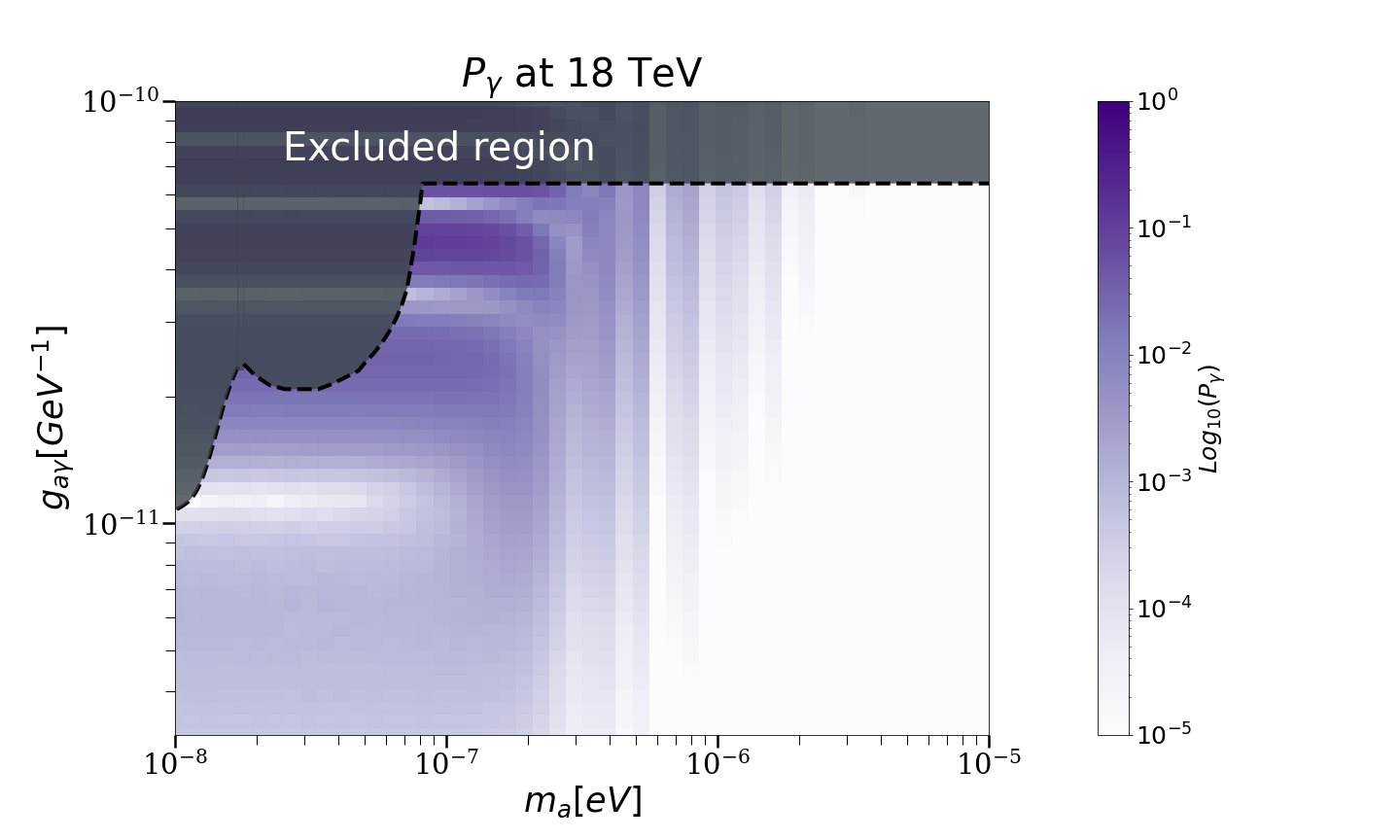}
\caption{Survival probability of an 18 TeV-photon from an initial beam of ALPs represented in a mesh of $m_{a}$ and $g_{a\gamma}$. The black dotted line represents the limits of the exclusion regions  that have been placed by Fermi \citep{https://doi.org/10.48550/arxiv.1304.0700}, H.E.S.S. \citep{Ajello_2016}, and CAST \citep{2017NatPh..13..584A}. The exclusion regions are located above the dotted line (gray shaded region). It can be seen that there are still candidates that could explain the detection of TeV-photons.}
\label{fig:probmesh}
\end{figure}

Different ALPs candidates, identified through the $m$ and $g_{a\gamma}$ parameters are considered. Figure \ref{fig:probcases} shows $P_{\gamma}$ for a $m=10^{-8} eV$, $g_{a\gamma}=10^{-10}GeV^{-1}$;  $m=10^{-7} eV$ , $g_{a\gamma}= \times 10^{-11} Gev^{-1}$; and $m=9 \times 10^{-7} eV $, $g_{a\gamma}=5\times 10^{ -12}GeV^{-1}$. Interesting candidates are those for which $P_{\gamma}$ is larger than the EBL attenuation at a given energy. For instance, the EBL attenuation factor is $0.18$, $1.6\times 10^{-7}$, and $4\times 10^{-77}$ for photon energies of 500 GeV, 18 TeV, and 250 TeV, respectively. Thus, we only show those candidates with survival probabilities larger than the minimum value to describe LHAASO observations obtained in subsection \ref{sec:lhaaso}. As observed, depending on the candidate, we could have photon contributions at different energy ranges. For instance, the candidate shown as a blue solid line will contribute with photons at the three energies of 0.5, 18, and 251 TeV. This could explain also the extremely bright observation by LHAASO at 500 GeV while the other two candidates will not. However, in this case, the first candidate is in the excluded region by the collaborations of FERMI, H.E.S.S., and CAST (the region above the dashed line of Figure \ref{fig:probmesh}) \citep{https://doi.org/10.48550/arxiv.1304.0700, Ajello_2016, 2017NatPh..13..584A}. More studies must be carried out to show if there exists a candidate with this behavior in the region not excluded. Candidates like the one shown in the dotted blue line are of special interest to explain the 251 TeV-photon since the survival probability increases with the photon energy. For the third candidate shown as a dashed blue line, there is an energy range between 18 and 251 TeV where observations of photons will be more difficult as the survival probability decreases. As observed, the magnetic field of the GRB jet lowers the survival probability but at the energies of interest, it has almost no effect. 

We calculate the survival probability for a mesh of values for $m_{a}$ and $g_{a\gamma}$ for photons with energies of 18 and 251 TeV. Figure \ref{fig:probmesh} shows those candidates with a survival probability above $10^{-5}$ for a 18 TeV photon, the plot for the region of a 251 TeV photon is omitted as it is contained within the 18 TeV photon region. The candidates above the dashed line are already excluded by Fermi or H.E.S.S. As observed, there exists a vast number of not excluded candidates that could explain the observation of 18 or 251 TeV photons. Note that this region has a strong energy dependency. Those candidates closer to the excluded region with lower masses would be more likely to explain both observations simultaneously.

\subsection{Dark photons}

Several conversion probabilities between photons and dark photons have previously been calculated in the literature. These expressions depend on the experimental setup envisioned to constrain the parameters of the dark photons, see for example \citep{Okun,paraphotons,hpfunk,wispydm,crabhp}, and \citep{Raffelt:1987im} for a general expression of interactions between photons and light particles and their effects. On top of that, in previous years, proposals to use astrophysical sources to observe a possible effect of the oscillation of photons to dark photons were suggested. For example in \citep{Okun}, it is proposed to observe if photons from a star can reach Earth when the star is eclipsed by the moon. Here, we use the method presented in \citep{crabhp} to obtain allowed values of the mixing angle $\chi$ and the mass of the dark photon $m_{\gamma\prime}$ to explain the TeV-photons observed from GRB 221009A. We assume the optical properties of the medium traversed by the photon are such that the refractive index is $1$\footnote{the refractive index is computed from the plasma frequency $\nu_p$, related to the electron density $n_e$ in the medium, as $\sqrt{1-\nu_p^2/\nu_{\gamma}}$. For $n_e\thicksim0.03~\text{cm}^{-3}$, and TeV-photons ($\nu_{\gamma}\gg\nu_p$) the refractive index is $1$.}, and that the absorption coefficient can be neglected for photons with energies above $100~\text{GeV}$. Under these assumptions, the expression for the conversion probability is given by \citep{paraphotons,crabhp}:

\begin{equation}
    P_{\gamma\to\gamma\prime} = 4\chi^{2}\sin^2\left(\frac{m_{\gamma\prime}^2L}{4E_\gamma}\right)
    \label{ptohp}
\end{equation}

where $L$ is the distance traveled by the photon and $E_\gamma$ is the photon energy. This expression corresponds to the case where the external magnetic field is zero which is not our case since we have to consider the magnetic field of the host galaxy, the Milky Way, and the GRB jet. \citet{Fortin_2019} gives an expression of the conversion probability when the external magnetic field is not zero but valid only for photons with energies of $E_\gamma\thicksim~\text{MeV}$. Thus, we take Equation \ref{ptohp} as a first approximation. The oscillation length is given by,

\begin{equation}
    L_{\text{osc}} = 2.56\times10^{-2}\left(\frac{E_\gamma}{1~\text{TeV}}\right)\left(\frac{m_{\gamma\prime}}{1~\mu\text{eV}}\right)^{-2} ~\text{pc}.
    \label{losc}
\end{equation}

\begin{figure}[ht!]\centering   
\epsscale{2}
\includegraphics[width=0.8\textwidth]{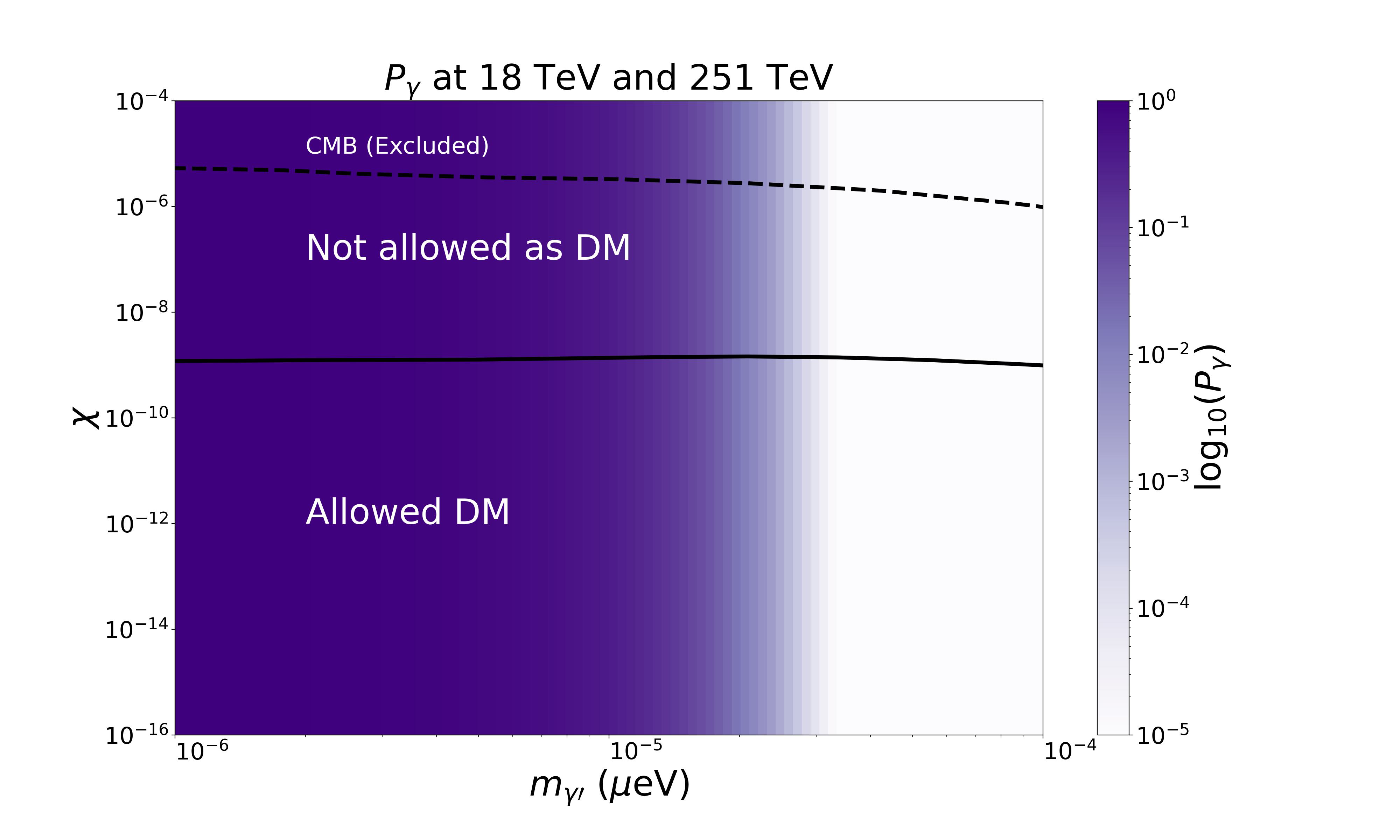}
\caption{Survival probability of photons with energies of $18~\text{TeV}$ or $251~\text{TeV}$ for dark photons with masses $m_{\gamma\prime}$ and mixing angle $\chi$. The region below the solid line shows the region where dark photons are a viable CDM candidate (Allowed DM). The region above the dashed line is excluded by using observations from the CMB (CMB excluded). The region between the solid and dashed lines is excluded as dark photons can not explain DM. The data for the allowed and excluded regions are taken from \citep{wispydm}.}
\label{survphpmesh}
\end{figure}

For energies $\thicksim10~\text{TeV}$, the oscillation length is smaller than the EBL attenuation length estimated between $2.4$ and $3.4~\text{Mpc}$ \citep{Schlickeiser_2012,neronov}. Additionally, we consider the EBL attenuation of the TeV-photons for the effective distance traveled by the photon. The final expression for the survival probability is given by,

\begin{equation}
    P_{\gamma} = (1-P_{\gamma\to\gamma\prime})\times \exp\left(-\tau(E_{\gamma},z)\frac{D_L}{2L_{\text{osc}}}\right)
    \label{survphp}
\end{equation}

\noindent
where $\tau(E_\gamma,z)$ is the optical depth as a function of the photon energy $E_{\gamma}$, and the redshift $z$. The luminous distance $D_L$ is given in Table \ref{tab:param}, and the oscillation length $L_{\text{osc}}$ is given by equation \ref{losc}. We use the public code \texttt{ebl-table} \citep{ebltable} to estimate the EBL attenuation. We calculate the survival probability for dark photon masses, $m_{\gamma\prime}$, in the range from $10^{-6}$ to $10^{-4}~\mu\text{eV}$, and the mixing angle $\chi$ with values between $10^{-6}$ and $10^{-4}$. Figure \ref{survphpmesh} shows the parameter space of 
$m_{\gamma\prime}$ and $\chi$ when $P_{\gamma}>10^{-5}$. In this case, the survival probability is independent of the observed photon energy. We observe that dark photons with masses smaller than $\thicksim3\times10^{-5}~\mu\text{eV}$ can explain the TeV-photons observed by LHAASO-KM2A. We also find that the survival probability $P_{\gamma}$ does not depend on the value of the mixing angle $\chi$. The permitted region for dark photons to be considered as DM capable to explain the observations from LHAASO and CARPET 2 are where the survival probability is required to be larger than $10^{-5}$ and outside of the excluded region. A strong difference with the ALP scenario is that dark photons will either explain simultaneously both 18 and 251 TeV-photons or neither of them.

%%%%%%%%%%%%%%%%%%%%%%%%%%%%%%%%%%%%%%%%%%%%%%%%%%%%%%%%%%%%%%%%%%%%%%%%%%%%%%%%%%%%%%%%%%%%%%%%%%%%%%%%%%
%%%%%%%%%%%%%%%%%%%%%%%%%%%%%%%%%%%%%%%%%%%%%%%%%%%%%%%%%%%%%%%%%%%%%%%%%%%%%%%%%%%%%%%%%%%%%%%%%%%%%%%%%%
\subsection{Previous GRB observations at energies \texorpdfstring{$ > 100$}{} GeV}\label{sec:past}

GRB observations have been one of the major challenges for the current VHE $\gamma$-ray astronomy observatories. In particular, imaging atmospheric Cherenkov telescopes (IACTs) have implemented observational programs to search for VHE emissions from GRBs.
Most results have provided upper limits to the photon flux and constraints to the models. However, MAGIC and H.E.S.S. are the only two IACTs that have detected GRBs at energies above $100\;\mathrm{GeV}$ \citep{2019Natur.575..455M,2019Natur.575..464A,2021Sci...372.1081H,201216C..MAGIC}. Table \ref{tab:PrevObs} resumes some relevant information about these observations. A common characteristic along these VHE GRBs is to be classified as long GRBs ($T_{90}\geq 30 \;\mathrm{s}$), except for GRB190829A that, when analyzing its energy flux and luminosity light curve, shows a rather different behavior \citep{Noda:2022hbo}. It is worth recalling that this sample contains only four bursts of several thousand observed bursts, so it may not be representative and no significant conclusions  should arise from it. Nevertheless, we can explore if DM can also explain the observations at energies of hundreds of GeV.

\begin{table}[ht!]
\centering
\begin{tabular}{cccccccc}
\hline
\multicolumn{8}{c}{GRB detections at VHE} \\
\hline
Instrument & GRB  & Energy & Normalization  & Pivot Energy & Spectral & Redshift & E$_{iso}$  \\
 &   & Range & [ TeV$^{-1}$ cm$^{-2}$ s$^{-1}$]  & [TeV] & Index &  & [erg]   \\
\hline
H.E.S.S. & 180720B & $100-440$ GeV & $7.52\times10^{-10}$ & 0.154  & $1.6\pm 1.2$ & 0.654 & $6.0\times10^{53}$ \\
MAGIC & 190114C & $300$ GeV $-$ $1$ TeV & $1.74\times10^{-7}$ & 0.386 & $2.16_{-0.31}^{+0.29}$ & $ 0.4245$ & $2.5\times10^{53}$ \\
H.E.S.S.& 190829A & $200$ GeV $-$ $4$ TeV & $-$ & $-$ & $-$ &0.0785 &$\sim 2\times 10^{50}$ \\
MAGIC & 201216C & $50-200$ GeV & $-$ & $-$ & $-$ & $ 1.1$ & $4.71\times10^{53}$ \\
\hline
\end{tabular}
\caption{Previous observation of GRBs at energies above $100$ GeV performed by IACTs}
\label{tab:PrevObs}
\end{table}

\begin{figure}[ht]\centering   
\epsscale{2}
\includegraphics[width=0.8\textwidth]{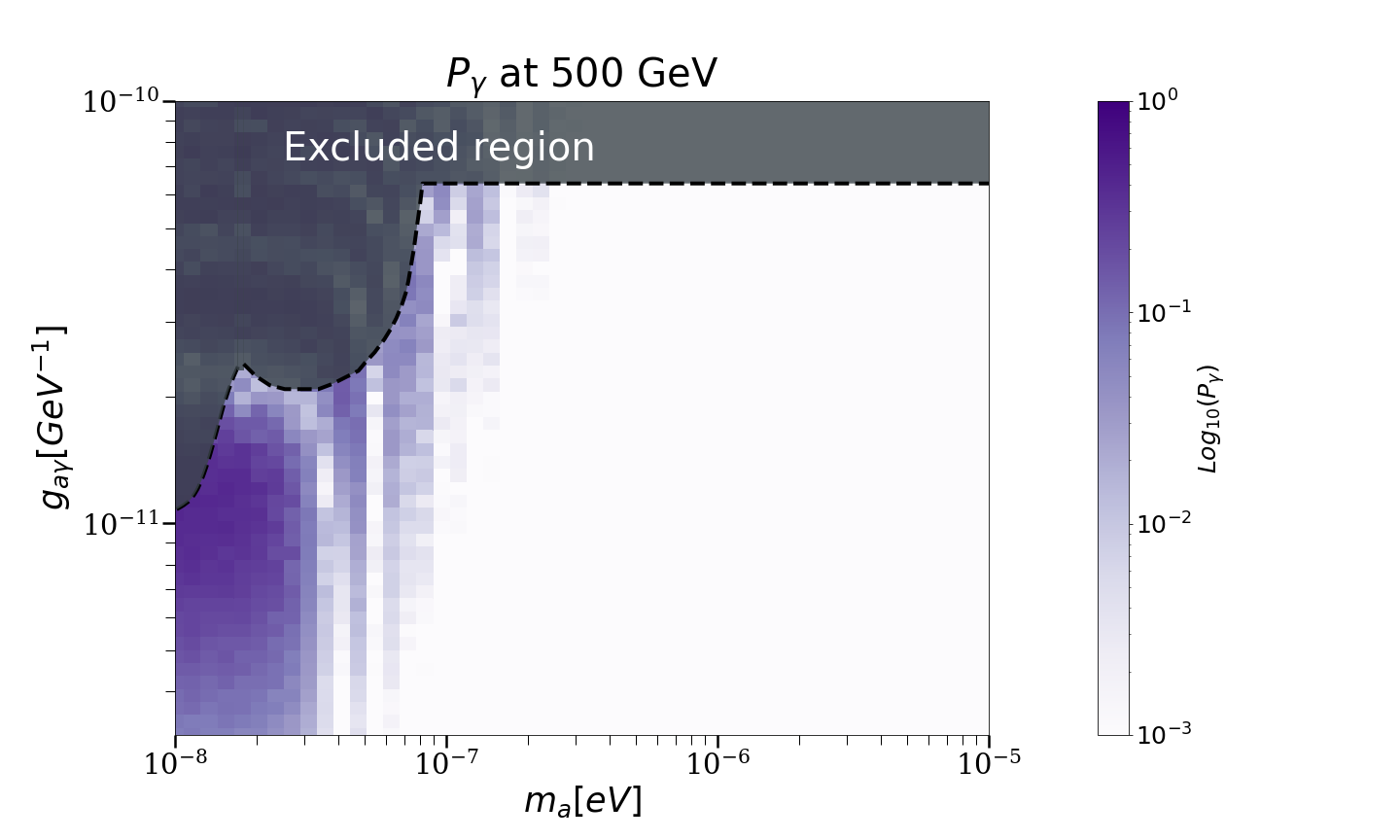}
\caption{Same as Figure \ref{fig:probmesh} but for photons with energy of 500 GeV.}
\label{fig:probmesh500}
\end{figure}

We have carried out the same procedure as in Section \ref{sec:lhaaso} to obtain the probability of DM particles to survive as a photon according to the reported fluxes in Table \ref{tab:PrevObs} after attenuation by the EBL.
Again, we consider a load energy of 1 and 10$\%$ of $E_{iso}$ corresponding to each burst as well as other parameters as the redshift given also in Table \ref{tab:PrevObs}. Table \ref{tab:Norm} shows the calculated survival probabilities. They are remarkably similar to each other considering the flux differences for the same energy fraction. As observed, larger values of the survival probability than those required for TeV-photons are also calculated. For the case of ALPs, we show in Figure \ref{fig:probmesh500} the candidate space for survival probabilities higher than $10^{-3}$. It is observed that there are still candidates out of the exclusion region but with low values of the coupling constant and the lowest masses. In the case of dark photons, the candidate region is slightly pushed to lower masses as seen in Figure \ref{survphpmesh}. Interestingly, these results propose DM as an alternative scenario to describe the observations of photons with energies of hundreds of GRBs. 

\begin{table}[ht!]
\centering
\begin{tabular}{cccc}
\hline
GRB name & $E_{\rm iso}$ & $P_{\gamma}$ & Instrument \\
\hline
190114C  & $0.01E_{\rm iso}$ & $5.1\times10^{-2}$ & MAGIC\\
180720B  & $0.01E_{\rm iso}$ & $1.1\times10^{-2}$ & H.E.S.S.\\
\hline
190114C  & $0.1E_{\rm iso}$ &  $5.1\times10^{-3}$ & MAGIC\\ 
180720B  & $0.1E_{\rm iso}$ &  $1.1\times10^{-3}$ & H.E.S.S \\
\hline
\end{tabular}
\caption{Calculated survival probability for photons with energies above 100 GeV for a given assumed fraction of $E_{iso}$ taken by DM particles.}
\label{tab:Norm}
\end{table}

%%%%%%%%%%%%%%%%%%%%%%%%%%%%%%%%%%%%%%%%%%%%%%%%%%%%%%%%%%%%%%%%%%%%%%%%%%%%%%%%%%%%%%%%%%%%%%%%%%%%%%%%%%
%%%%%%%%%%%%%%%%%%%%%%%%%%%%%%%%%%%%%%%%%%%%%%%%%%%%%%%%%%%%%%%%%%%%%%%%%%%%%%%%%%%%%%%%%%%%%%%%%%%%%%%%%%
%%%%%%%%%%%%%%%%%%%%%%%%%%%%%%%%%%%%%%%%%%%%%%%%%%%%%%%%%%%%%%%%%%%%%%%%%%%%%%%%%%%%%%%%%%%%%%%%%%%%%%%%%%
%%%%%%%%%%%%%%%%%%%%%%%%%%%%%%%%%%%%%%%%%%%%%%%%%%%%%%%%%%%%%%%%%%%%%%%%%%%%%%%%%%%%%%%%%%%%%%%%%%%%%%%%%%
\section{Discussion and Conclusions}  \label{sec:disc}

We have investigated three scenarios. First, we take the most plausible mechanism within the fireball scenario, the SSC emission in the external forward-shock model, based on the long duration of the VHE gamma-ray emission and the 18-TeV photon observed. We found that to avoid the Klein-Nishina regime, very small and unlikely values of the microphysical parameters are required, and even before considering the EBL attenuation; the resulting flux is too small to reproduce the observation of at least one photon with an energy of 18 TeV. \citet{2022arXiv221010673R} modeled the multiwavelength afterglow observations of GRB 221009A in a stellar-wind-dominated environment considering the Klein Nishina effect and find its emission in the energy range of 0.1 to 10 TeV to be very bright, peaking at 300 GeV. Even though the calculated flux at 1 TeV in the first thousand seconds, $\sim 10^{-9}$ ${\rm erg\, cm^{-2}\, s^{-1}}$, is below the HAWC upper limit, final calculations need to be performed to assure that the LHAASO observations can be reproduced since the corresponding EBL attenuation is $\sim 10^{-7}$ at energies of 18 TeV. The forthcoming LHAASO light curve and spectra for GRB 221009A may provide conclusive information about whether or not SSC is the appropriate mechanism, although with the Klein Nishina break energy below 1 TeV we consider it unlikely and impossible for explaining the 251 TeV-photon. Therefore, it is not clear that a population of TeV-photons is truly released in this burst at least by SSC emission alone, leading us to explore other alternatives involving DM.

Then, we consider two DM scenarios that can be coupled with TeV-photons, ALPs, and dark photons. Several authors, such as \citet{https://doi.org/10.48550/arxiv.2210.05659}, \citet{https://doi.org/10.48550/arxiv.2210.07172}, and \citet{https://doi.org/10.48550/arxiv.2210.09250} have considered a starting population of photons that are transformed into DM and in the MW is where they get reconverted to photons. As discussed before, we believe that the release of TeV-photons from GRB221009A is very unlikely then instead, we start with a beam of light DM particles released by the burst. How these DM particles are released, created, or accelerated is out of the scope of this paper but new mechanisms have emerged such as the one described in \cite{https://doi.org/10.48550/arxiv.2210.10022}. In this study, the production of ALPs from first-order phase transitions is considered to explain the observation of VHE photons from GRB 221009A. Interestingly, the region of viable parameters within this mechanism includes the region of parameters that we find here to explain the 18 TeV photon. Nonetheless, we have pointed out theories that consider scenarios of DM production in GRBs, which still need to devise a mechanism to better explain the acceleration of DM particles up to TeV energies. Most of these theories identified the progenitor as the site with the required conditions to generate DM. We assume that the DM release happens very close to the trigger time and lasts over $2\times10^{3}~\text{s}$. Nevertheless, this is not necessarily true and may contradict models where the progenitor produces the DM as it evolves \citep{alpSNI,alpSNII}. 

We find DM candidates outside the excluded regions which are capable of explaining the observation of TeV photons in GRB 221009A. In the case of ALPs, candidates with lower masses and higher coupling coefficients could explain both, 18 and 251 TeV photons simultaneously; however, they are close to the excluded region. In the case of dark photons, an explanation of both photons comes naturally at least when the survival probability does not depend on the magnetic field. Under this hypothesis, the TeV-photons would acquire spectral properties from dark photons.  But more detailed studies are needed to estimate the dependence of the survival probability with the magnetic field.

We have shown how the ALP's scenario could or could not contribute at different energy ranges. In particular, we have explored the parameter space to describe previous observations of GRBs at hundreds of GeV as well as the LHAASO observation of GRB 221009A. Again, this may come naturally if there exist light-dark photons with energies of hundreds of GeVs while in the case of ALPs the allowed region decreases dramatically towards the lowest masses and low coupling constants. The values found of the survival probability leave the door open to ALPs to explain hundred of GeV emission. Nevertheless, a detailed study considering different DM spectra, including spectral lines, is required. The confrontation of the results with the spectra observed beyond 10 TeV will definitely help to distinguish the mechanism necessary to produce such a high energy emission from very distant sources.

We have calculated survival probabilities for our two DM scenarios considering the burst's observable parameters. For those unknown quantities, we have used conservative values for long bursts. We have included the jet, host galaxy, and Milky Way environments. We have found a loss of $\sim 30\%$ of photons by conversion of ALPs before reaching the Milky Way. This is understood as a consequence of the small jet size and the low magnetic field of the host galaxy. The correction by considering the intergalactic magnetic field is neglectable. For dark photons, a detailed estimation of the dependency of the survival probability with the magnetic field must be carried out if they remain as candidates. However, we have shown that dark photons could also be possible candidates even though ALPs have generated a lot more studies. Our assumptions on the energy taken by DM from the burst are conservative and plausible when compared with the fraction of energy taken as kinetic energy of electrons responsible for the emission in other lower frequencies. Thus, the introduction of DM would not affect other elements of the burst evolution. If DM is present in GRBs, the information carried by the light curve will be crucial in order to disentangle possible theories to explain the production and release mechanisms of DM in GRBs, as well as the nature of the DM particle involved.

%\section*{Acknowledgments}

This work was supported by UNAM-PAPIIT project numbers IG101320 and IG101323.

\bibliography{sample631}{}
\bibliographystyle{aasjournal}



\end{document}